\documentclass[acmtog, nonacm]{acmart}
\acmSubmissionID{364}
\usepackage{subcaption}
\usepackage{hyphenat}

\renewcommand\footnotetextcopyrightpermission[1]{} 
\usepackage[ruled]{algorithm2e} 
\usepackage{xr} 
\usepackage{multicol}

\SetAlFnt{\small}
\SetAlCapFnt{\small}
\SetAlCapNameFnt{\small}
\SetAlCapHSkip{0pt}

\acmJournal{TOG}

\copyrightyear{2023} 
\acmYear{2023} 
\setcopyright{rightsretained} 
\acmDOI{XX.XXXX/XXXXXXX.XXXXXXX}

\begin{document}
	
	\title{Saccade-Contingent Rendering}
	\author{Yuna Kwak}
	\affiliation{
			\institution{Reality Labs Research, Meta and New York University}
			\country{USA}
		}
			\email{yunakwak@meta.com}
			
	\author{Eric Penner}
	\affiliation{
			\institution{Reality Labs Research, Meta}
			\country{USA}
		}
	\email{epenner@meta.com}
	
	\author{Xuan Wang}
	\affiliation{
			\institution{Reality Labs Research, Meta}
			\country{USA}
		}
	\email{xuanwang0925@meta.com}

	\author{Mohammad R. Saeedpour-Parizi}
	\affiliation{
			\institution{Reality Labs, Meta}
			\country{USA}
		}
	\email{rezasaeedpour@meta.com}

	\author{Olivier Mercier}
	\affiliation{
			\institution{Reality Labs Research, Meta}
			\country{USA}
		}
	\email{omercier@meta.com}

	\author{Xiuyun Wu}
	\affiliation{
			\institution{Reality Labs, Meta}
			\country{USA}
		}
	\email{xiuyunwu@meta.com}
	
	\author{T. Scott Murdison}
	\affiliation{
			\institution{Reality Labs, Meta}
			\country{USA}
		}
	\email{smurdison@meta.com}

	\author{Phillip Guan}
	\affiliation{
			\institution{Reality Labs Research, Meta}
			\country{USA}
		}
	\email{philguan@meta.com}
	
	\begin{abstract}
        Battery-constrained power consumption, compute limitations, and high frame rate requirements in head-mounted displays present unique challenges in the drive to present increasingly immersive and comfortable imagery in virtual reality. However, humans are not equally sensitive to all regions of the visual field, and perceptually-optimized rendering techniques are increasingly utilized to address these bottlenecks. Many of these techniques are gaze-contingent and often render reduced detail away from a user's fixation. Such techniques are dependent on spatio-temporally-accurate gaze tracking and can result in obvious visual artifacts when eye tracking is inaccurate. In this work we present a gaze-contingent rendering technique which only requires saccade detection, bypassing the need for highly-accurate eye tracking. In our first experiment, we show that visual acuity is reduced for several hundred milliseconds after a saccade. In our second experiment, we use these results to reduce the rendered image resolution after saccades in a controlled psychophysical setup, and find that observers cannot discriminate between saccade-contingent reduced-resolution rendering and full-resolution rendering. Finally, in our third experiment, we introduce a 90 pixels per degree headset and validate our saccade-contingent rendering method under typical VR viewing conditions.
	\end{abstract}
	

	
	

	\begin{teaserfigure}
		\centering
		\includegraphics[width=\textwidth]{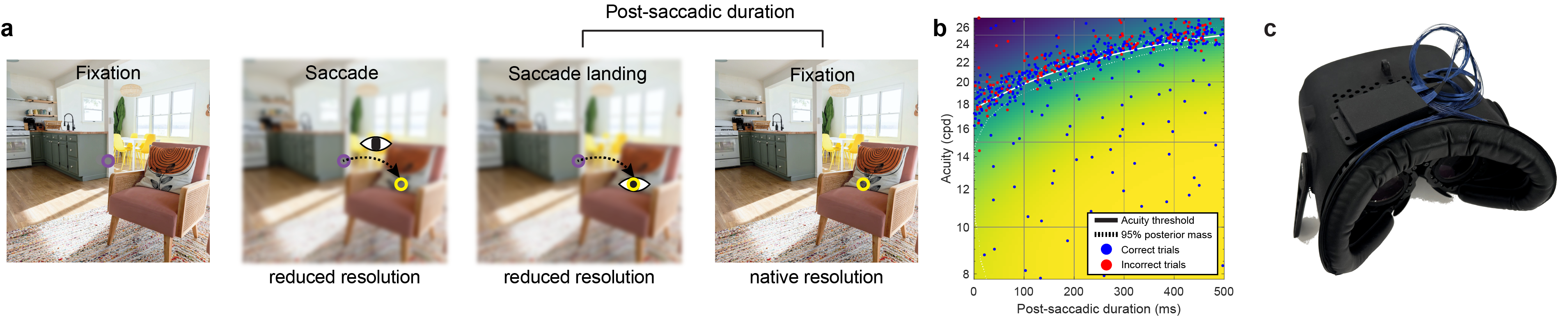}
 		\caption{We propose a novel saccade-contingent rendering algorithm which takes advantage of reduced visual acuity immediately after rapid shifts in gaze (i.e., saccades). (a) Our algorithm reduces the rendered image resolution after a detected saccade and then increases the render resolution over several hundred milliseconds (post-saccadic duration). The algorithm is informed and validated across three user studies. (b) In the first study we measure the temporal evolution of post-saccadic visual acuity with a low-level psychophysical stimulus. Here, one participant's data is presented, showing how visual acuity improves during the 500 ms following saccade completion (i.e., saccade landing). Blue regions indicate estimated chance (50\% correct) task performance and yellow indicates estimated 100\% correct task performance. Spatial acuity is defined as the 75\% correct performance. (c) In the third and final experiment, we implement our algorithm on a 90 pixel per degree headset to evaluate it with real VR content and more natural head and eye movements.}
		\label{fig:teaser}
	\end{teaserfigure}
	\maketitle
	
	\section{Introduction}
\label{sec:intro}

Perceptually-optimized compression techniques are routinely used to address rendering, power, bandwidth, storage, and cost constraints while minimizing impacts to perceived image quality across a variety of applications in traditional direct-view displays (e.g., interlacing in TV, jpg compression in photos). Head-mounted displays (HMDs) introduce additional computational challenges in all of these areas; retinal resolution, wide field of view virtual reality (VR) headsets require rendering many pixels at high frame rates (>90 Hz) in mobile form factors with limited compute and power envelopes.


The human visual system is not equally sensitive to visual input over its field of view, and gaze-contingent foveated rendering was proposed to take advantage of this fact in head-mounted displays (HMDs) \cite{guenter2012foveated, PatneyEtAl2016}. Such methods are able to maintain a user's perception of a full-fidelity scene while rendering at lower quality levels in the user's peripheral vision \cite{WatsonEtAl1996effectiveness,WatsonEtAl1997detail,SwaffordEtAl2016}, but these methods require spatially-accurate eye tracking. Eye trackers used for foveated rendering must also have low latency to account for user saccades, i.e., rapid shifts to new fixation points during natural gaze \cite{AlbertEtAl2017, arabadzhiyska2017saccade}.

Saccades allow humans to compensate for lower visual acuity in the periphery by placing features of interest at the fovea, where spatial acuity and color perception are most sensitive \cite{virsu1979visual, bowmaker1980visual}. Visual input during a saccade is suppressed to hide retinal motion blur, a widely-known phenomenon called saccadic suppression \cite{idrees2020perceptual, niemeier2003optimal, Burr1994,RossEtAl2001,Matin1974saccadic}. Notably, visual sensitivity does not immediately recover upon completion of the saccade \cite{diamond2000extraretinal}. Saccades to a new gaze point are followed by smooth fixational motions called ocular drifts as well as rapid, small saccades ("microsaccades")~\cite{RucciPoletti2015}, and these involuntary eye movements increase visual sensitivity to high spatial frequencies over time \cite{clark2022eye, CoxEtAl2022, KuangEtAl2012}. In other words, foveal visual acuity immediately following saccades is \textit{reduced} until the eyes can make ocular drifts and microsaccades.

While spatial variations in visual perception have been exploited for foveated-rendering techniques, none have taken advantage of this temporal in acuity change following the completion of a saccade (i.e., the post-saccadic landing period). We propose a novel, \textit{non\hyp{}foveated} saccade\hyp{}contingent rendering algorithm which leverages this phenomenon. We validate this approach with three user studies and make these additional contributions:

\begin{itemize}{}
    \item For the first time, we fully characterize post-saccadic visual acuity across time and spatial frequency. Using a low-level psychophysical task, we identify significantly reduced spatial acuity after saccade completion followed by a rapid increase over 100-200 milliseconds.
    
    \item Based on these newly identified perceptual data, we show that saccade-contingent rendering can yield >50\% bandwidth savings on high\hyp{}resolution displays (>60 pixels per degree).
    
    \item We evaluate saccade-contingent rendering in a controlled and replicable psychophysical experiment across 30 natural images and confirm the algorithm's viability.
    
    \item Finally, we build a 90 pixel per degree (ppd), eye-tracked VR headset. We use this headset to further evaluate saccade\hyp{}contingent rendering during ecologically-valid, unconstrained VR viewing conditions.
    
\end{itemize}
	\section{Related Work}
\label{sec:related_work}

\paragraph{Saccades in Human Vision}
Three to five saccadic eye movements are made every second \cite{kowler2011, fabius2019time,gray2014adaptation}, each typically lasting less than 50 milliseconds while reaching speeds in excess of 200$^\circ$ per second \cite{robinson1964mechanics,land1999roles}. Saccade amplitudes and velocities typically follow a lawful kinematic relationship called the saccadic main sequence \cite{bahill1975most}. This deterministic saccadic behavior creates predictable retinal blur, which the visual system combines with retinal flow during fixational periods to form a cohesive and stable perception of space -- a process that underscores the importance of temporal integration in spatial perception \cite{KuangEtAl2012,BoiEtAl2012,MostofiEtAl2020,schweitzer2020intra,rolfs2022coupling,kroell2022foveal,murdison2019saccade,goettker2020differences}. In Experiment~1 we measure the temporal evolution of visual acuity following a saccade landing and explore the potential savings from saccade-contingent rendering  based on natural saccade behavior.




\paragraph{Spatiotemporal Vision Models}
Early contrast-sensitivity measurements of human vision only account for spatiotemporal sensitivity for a limited range of stimuli \cite{Robson1966, Kelly1979, koenderink1978visual}. Perceptual modeling based on these data is similarly limited and can only account for basic visual stimuli and foveal viewing \cite{watson1986window}. Subsequent extensions to these models introduce additional viewing contexts and incorporate factors such as luminance, eccentricity, and color \cite{mantiuk-hdrvdp2, FovVideoVDP, stelaCSF, TursunEtAl2019, tursun2022perceptual, WatsonAhmuda2016}. Few models account for changes in perception with active observers making head and eye movements. Daly~\shortcite{Daly2001} introduced a model which incorporates some effects of eye movements, albeit on a longer time scale which does not account for the dynamic changes in intra- and post-saccadic visual acuity, the latter of which we measure in Experiment~1.

\paragraph{Foveated Rendering in HMDs}
Most perceptual optimizations in HMDs exploit eccentricity\hyp{}dependent limitations of the human visual system \cite{matthews2020rendering}. Guenter et al. \shortcite{guenter2012foveated} and Patney et al. \shortcite{PatneyEtAl2016} demonstrate the first real-time foveated rendering based on psychophysical studies modeling visual acuity falloffs with eccentricity. Recent work in this domain typically expands upon existing spatiotemporal visual models to incorporate additional perceptual optimizations such as color \cite{duinkharjav2022color}, content-dependent foveation \cite{TursunEtAl2019, tariq2022noise}, peripheral spatiotemporal sensitivity \cite{tursun2022perceptual}, and attention \cite{krajancich2023towards}. These foveation techniques rely on accurate gaze tracking, and the rapid, ballistic nature of saccades can introduce visible artifacts depending on eye tracking latency. Albert et al.~\shortcite{AlbertEtAl2017} measured latency requirements for different foveation techniques and conclude that an eye-to-photon latency between 50-70 ms (or better) is required for many of them, which can be difficult to achieve in practice. In this work we propose the first perceptually\hyp{}optimized, gaze\hyp{}contingent compression technique that is non-foveated and rigorously confirm its viability in Experiment~2. We further validate its application in real hardware, introducing a 90 ppd HMD with 30 ms eye to photon latency in Experiment~3. 

\paragraph{Non-Foveated Perceptual Rendering Optimizations in HMDs}
A smaller number of non-foveated perceptual optimizations for rendering in HMDs are available. Binocular summation can be leveraged to reduce frame rate in one eye \cite{denes2019temporal} and spatiotemporal models of vision can be used to tradeoff between refresh rate and render resolution \cite{denes2020perceptual}. Spatiotemporal vision models can also be used to apply local changes in render quality \cite{jindal2021perceptual} without foveation.  Saccade-contingent rendering, however, is distinct from these previous proposals. We show that saccades introduce brief temporal reductions in spatial contrast sensitivity, a behavior that is not captured in existing spatiotemporal contrast sensitivity models underlying these foveated and non-foveated spatiotemporal optimizations. Our proposed algorithm identifies a unique, and preciously un-leveraged phenomenon in human perception and can be used \textit{in addition} to existing perceptually-optimized HMD rendering paradigms.

	\begin{figure*}[tb]
	\includegraphics[width=\textwidth]{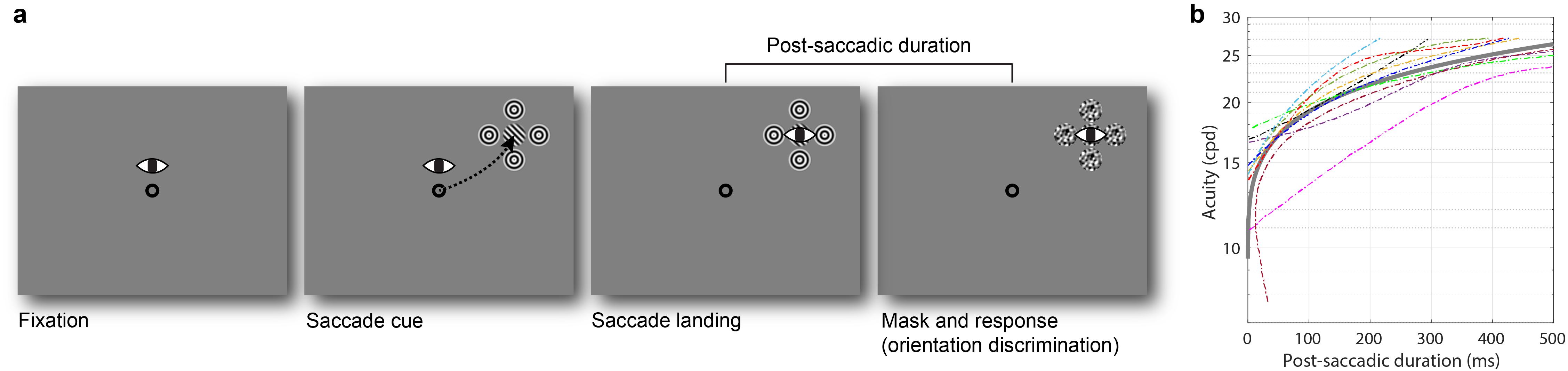}
	\caption{(a) Stimulus presentation for Experiment 1. Participant spatial acuity is determined as a function of post-saccadic duration with a Gabor filter orientation-identification task. (b) Acuity thresholds as a function of time after saccade landing. Colored lines show individual threshold curves, an example data set used to derive each participants' curve is shown in Figure~\ref{fig:teaser}b. Data for all participants can be found in the supplementary materials. The thick gray line denotes the power fit to the median of all participants' acuity thresholds (R$^2$=0.71).}
	\label{fig:exp1_results}
\end{figure*}

\section{Experiment 1: Psychophysical Study} \label{sec:exp1}
While spatial acuity is reduced during the post-saccadic landing period \cite{clark2022eye,diamond2000extraretinal}, a saccade-contingent rendering algorithm must be implemented based on a continuous measure of perceptual sensitivity through time. If parameters used to reduce render resolution are too aggressive, users will perceive visible artifacts, similar to a foveated rendering algorithm which has poorly\hyp{}tuned parameters for eccentricity-based falloff. Previous work has only characterized time-dependent acuity changes for two spatial frequencies~\cite{CoxEtAl2022} and in this experiment we present a precise, low-level psychophysical task used to characterize this space in enough detail to support saccade-contingent rendering.

\subsection{User Study} \label{sec:exp1_procedure}
We designed a two alternative forced-choice (2AFC) discrimination task to measure visual acuity in the time period following the completion of a saccade. The protocol is shown in Figure~\ref{fig:exp2}a. Users began by fixating on a target at the center of a display. After a random duration between 500-600 ms, a saccade target appeared 10.3$^\circ$ to 15.5$^\circ$ along the NW, NE, SW, or SE diagonals away from fixation. The saccade target is a vertically oriented Gabor patch rotated counter-clockwise (-45$^\circ$) or clockwise (+45$^\circ$) from vertical, and was surrounded by four circular distractor Gabor patches. The crowding effect of the distractor Gabors reduces the visibility of the center Gabor if the user do not make an accurate saccade, thereby enforcing more precise eye movements. The Gabor patches are replaced with bandpass filtered noise masks in the same location after a pre-determined stimulus presentation duration. In each trial users were asked to identify the orientation of the target Gabor patch (counter-clockwise or clockwise). The spatial frequency and presentation duration of the target Gabor stimulus were adjusted parametrically  with AEPsych \cite{owen2021,guan22} (see Section 1.2 in supplementary materials) before each trial to measure visual acuity as a function of time after saccade landing (i.e., post-saccadic duration). For each presentation, duration peak spatial acuity was defined as the highest spatial frequency Gabor stimulus where its orientation was correctly identified 75\% of the time. We provided auditory feedback after each trial, similar to other studies~\cite{guan2023perceptual} to improve participant engagement without affecting reproducibility~\cite{Bach2016}. Additional stimulus details and criteria used for online and offline eye tracking analysis can be found in Section~1 of the Supplementary Materials.

\subsection{Participants}
Ten users (six males and four females, mean age=28.2) with normal or corrected-to-normal visual acuity of 20/20 participated in the study. A total of 600 trials with correct fixation and eye movements were collected for each participant; all study protocols were reviewed and approved by a third-party Institutional Review Board.

\subsection{Hardware} \label{sec:exp1_app}
A 48" LG OLED TV (OLED48C2PUA) with a 3,840 $\times$ 2,160 resolution was placed 100 cm from the user's eyes to achieve a 55.6$^{\circ}$ $\times $ 33.1$^{\circ}$ FOV. The refresh rate was set to 120 Hz. All stimuli were generated and presented using MATLAB (MathWorks, Natick, MA, USA) and Psychtoolbox \cite{brainard1997,pelli1997,kleiner2007s}. Gaze position was monitored using an EyeLink 1000 desktop eye tracker (SR Research, Osgoode, Ontario, Canada) set to a sampling rate of 1000 Hz.

\subsection{Results} \label{sec:acuity_equation}
Changes in user spatial acuity after saccade landing are shown in Figure~\ref{fig:exp1_results}b. Individual data and trial-by-trial responses can be found in our supplementary materials (Figure S2). Users are able to resolve $\approx$10 cpd Gabor patches immediately upon saccade landing and visual acuity increases up to $\approx$27 cpd approximately 500 ms later. This behavior is well-described by a power function fit to participants' acuity thresholds (R$^{2}$ = 0.71):

\begin{equation}  \label{eqn:sf_curve}
    sf(t) = 1.9469t^{0.3475} + 9.5062
\end{equation}
where $sf$ and $t$ denote the highest resolvable Gabor patch spatial frequency and the post-saccadic duration, respectively. With this description of post-saccadic visual acuity in place, we model the savings for our saccade-contingent rendering algorithm.

\begin{figure}[t]
	\includegraphics[width=.475\textwidth]{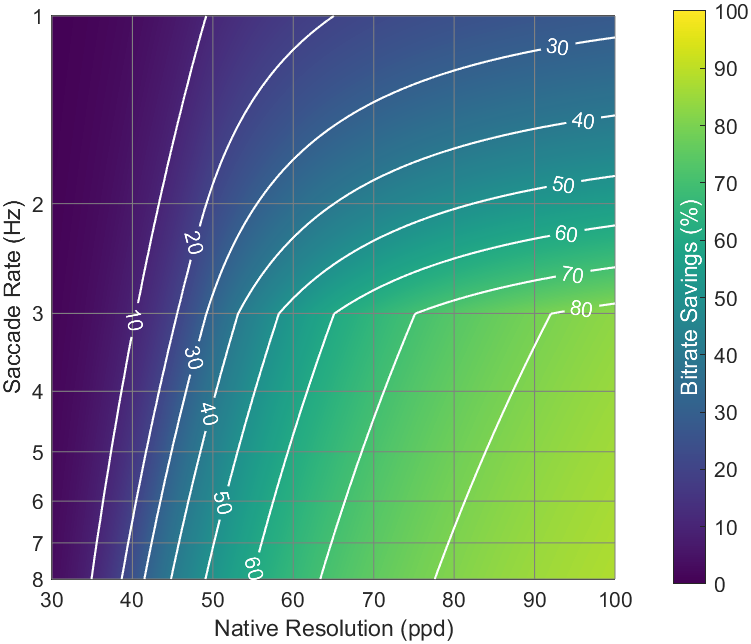}
    	\caption{Relative bitrate savings shown for a 90 Hz, 32-bit RGB display for 30-100 ppd native resolution displays and 1-8 Hz saccade frequencies. Higher saccade rates enable more-frequent saccade-contingent rendering and greatly increases bit savings as the most significant savings occur in the first 100~ms following saccade landing. Data collection from Figure~\ref{fig:exp1_results} is limited to 27 cpd and 500 ms, to account for these limitations in our simulation we revert to rendering at native-resolution after 333 ms. Most of the bit savings are achieved in the first 200 ms of our algorithm, so any cutoff >200 ms does not significantly influence this simulation $(\approx\pm20\%)$.}
    \label{fig:bandwidth}
\end{figure}

\subsection{Saccade-Contingent Rendering Bandwidth Analysis} \label{sec:bandwidth}
To our knowledge, this is the first study to document this temporal evolution in detail. Critically, visual acuity is highest at the fovea \cite{virsu1979visual}, with reduced post-saccadic visual acuity observed beyond the fovea \cite{li2021post}. This can be leveraged by rendering frames at lower resolution after saccades to enable power and compute savings. Unfortunately, it is difficult to directly model power or compute savings, as specific estimates are too closely tied to specific software and hardware architectures to be generalizable or meaningful. We instead examine bandwidth savings, which requires fewer assumptions and generalizes across architectures more easily, as an approximation for power and compute savings. We use Equation~\ref{eqn:sf_curve} to compute a first-order estimate of data savings in terms of bitrate (bits/s) across a 500 ms post-saccadic window for a 90 Hz, 32-bit RGB panel with an assumed display resolution of 3840 $\times$ 2160 to achieve 30 ppd. Since we present savings as a ratio of total bits, the absolute display resolution does not affect our analysis.

For a single display (i.e., for one eye), we compute bitrate savings as the ratio of bits for an image rendered at the resolution specified by Equation~\ref{eqn:sf_curve} and native display resolution. In our user study, post-saccadic duration is limited to 500 ms, but we note that the model in Equation~\ref{eqn:sf_curve} extends well beyond this window. For a more conservative measure of bit savings, we only simulate reduced-resolution rendering following Equation~\ref{eqn:sf_curve} for 333 ms, after which we simulate rendering at the native resolution of the display. Instantaneous bitrate savings (i.e., savings for a single frame at a given time) are highest immediately after a saccade -- when visual acuity is lowest and resolution reduction can be most aggressive. However, these savings are not significant for <35 ppd displays because they are only realized for a small number of frames before post-saccadic visual acuity quickly surpasses the native resolution of the display.

Beyond 35 ppd, the effective bit savings with saccade-contingent rendering depends strongly on the number of saccades users make, since the most significant savings are achieved in the first hundred milliseconds after each saccade landing. Higher saccade frequencies result in more frames rendered during the early period of the post-saccadic acuity curve. This relationship is highlighted in Figure~\ref{fig:bandwidth}, which shows how the relative bit savings changes based on saccade rate for 30-100 ppd displays (we assume saccades are uniformly distributed in time at each frequency). Using an average of three to five saccades from natural gaze statistics, we find modest bit savings at 40 ppd (9-16 percent), but observe rapidly-increasing savings at higher resolutions (32-44, 53-61, 65-71, and 73-78 percent at 50, 60, 70, 80 ppd, respectively). This bitrate simulation shows that significant savings can be achieved with saccade\hyp{}contingent rendering in HMDs if the headset's native resolution is 50 ppd or greater. In our next user study, we apply saccade-contingent rendering to real images to evaluate its feasibility with more natural content.

	\section{Experiment 2: Benchtop Validation}
Gabor stimuli only contain energy at a single spatial frequency, and it is important to confirm that the thresholds established in Experiment 1 apply to real images containing a wide mix of frequencies. In this experiment, we study the viability of saccade-contingent rendering across 30 images. However, there are a significant number of factors to consider for this evaluation, including the equation used to set post-saccadic resolution reduction, effects of image structure, individual variation in post-saccadic visual acuity, interactions with head and eye movements, and angular pixel density. We make a number of decisions to simplify this problem space so that we can evaluate whether saccade-contingent rendering can be applied without detection by users in a rigorous and repeatable manner.

\begin{figure*}[t]
	\includegraphics[width=.975\textwidth]{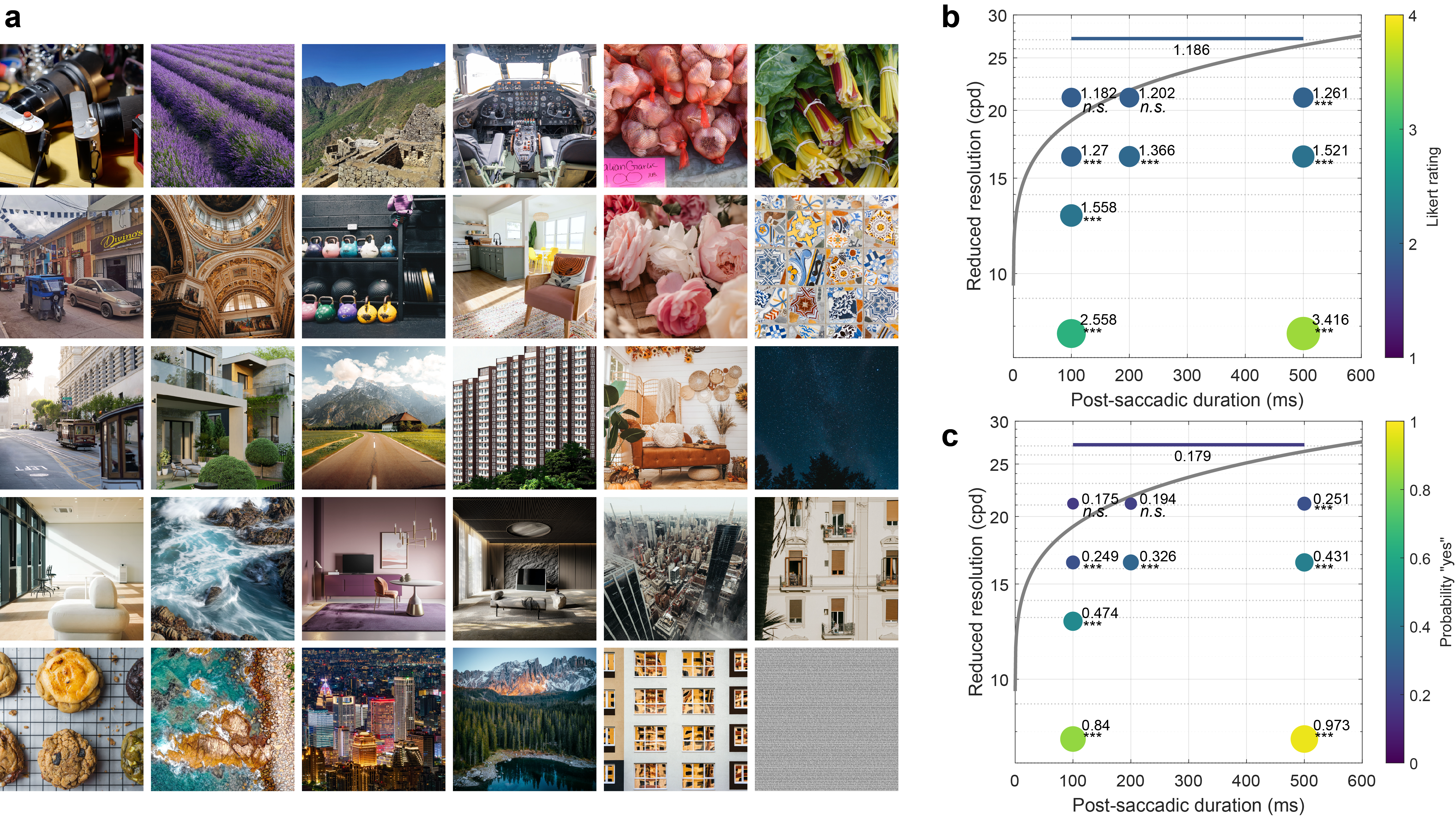}
	\caption{(a) Thirty different images used in Experiment~2. (b) Likert rating responses to how strong the resolution change was perceived, as a function of when the change occurred after saccade landing. Larger numbers indicate the change was perceived to a greater extent. (c) Likert rating responses analyzed into the probability of detecting changes. (b-c) Horizontal bars on the top indicate the baseline condition in which there was no change in rendered resolution. The color and size of the circles and bars reflect the Likert rating and probability numbers. The gray solid line indicates the group acuity thresholds from Experiment~1. Statistical significance from the baseline condition are denoted in asterisks. ***p<0.001, \textit{n.s.} not significant.}
	\label{fig:exp2}
\end{figure*}

\subsection{Experimental Design}
\paragraph{Downsampling Algorithm}
Saccade-contingent rendering is simulated by temporarily lowpass-filtering an image (additional details in Section 2.1 of supplementary materials). To eliminate additional complications in temporal tuning, we use a step function increase from post-saccadic render resolution back to native resolution. The step function approach is a conservative implementation of saccade-contingent rendering since a sudden change in image sharpness will be more apparent than a smooth transition. Five lowpass filter spatial frequencies (7.8, 12.8, 16.4, 21.1 cpd; ppd = cpd $\times$ 2) and three presentation times (100, 200, or 500 ms) were selected based on the acuity thresholds found in Experiment~1.

\paragraph{Image Content} Thirty different images (Figure~\ref{fig:exp2}a; sources available in Section~3 of the Supplementary Materials) are used in this experiment to control for the potential interactions of image content on saccade-contingent rendering. Ten combinations of step function-based saccade-contingent rendering are applied to each image, resulting in a total of 300 trials per participant (Figure~\ref{fig:exp2}).

\paragraph{Hardware}
We used the Eyelink 1000 and OLED display from Experiment 1 to ensure that eye tracking and display quality limitations did not introduce additional artifacts to our visual stimulus.

\paragraph{User Behavior}
We remove potential complications from differing head movements across participants by using a chin rest to stabilize all user's heads during the experiment. We control for eye movements by pre-determining a fixation point and providing a saccade target cue to initiate saccades to the same location for all observers (the saccade target is in a different location for each of our 30 images). These choices ensure that data collected in this study is comparable across all participants.


\subsection{User Study} \label{sec:exp2_procedure}
The experimental design is illustrated in Figure~\ref{fig:teaser}a. At the beginning of each trial, users fixated on a $0.4^\circ$ circle at the center of a high-resolution image (2000 $\times $ 2000 pixels) presented on a gray background (7.2 cd/m$^2$). After a random interval between 300-500 ms, a $6.4^\circ$ saccade target was presented in the periphery ranging from $3.6^\circ$ to $22.4^\circ$ (mean = $14.7^\circ$) away from fixation. After making a saccade and viewing the saccade-contingent rendering sequence, users rated perceived change in image clarity with a four-point Likert scale: 1 for no change, 2 for slight change, 3 for moderate change, and 4 for considerable change. There was also a baseline condition where image resolution was unchanged throughout the entire trial (\emph{no-change} condition). Blurred images are presented for 100, 200, or 500 ms after saccade landing before the non-blurred image is presented for another 400 ms. Some trials were excluded after offline analysis (Section~1.3 supplementary materials) if users did not make their saccades accurately. Before the main experiment, users were shown a version of the experiment with the same 10 combinations of low-pass filtering and image presentation times without making saccades. This was intended to anchor user responses with the Likert scale and to show users what changes in image quality to look for in the experiment. 

\subsection{Participants}
Thirty-six users (20 males and 16 females, mean age=34.8) with normal or corrected-to-normal visual acuity of 20/20 participated in the study. One user's data was excluded because the percentage of no-change responses (1 in the Likert scale) in the baseline (no-change) condition was only 6.7$\%$. Most of the responses in the baseline condition from this user were 2 ("slight change") and 3 ("moderate change"). All study protocols were approved by an external Institutional Review Board.

\subsection{Results}
Average Likert scale ratings for each combination of post-saccadic duration and low-pass spatial frequency cutoff over all 30 images is shown in Figure~\ref{fig:exp2}b. As expected, longer post-saccadic durations and more aggressive low-pass filtering result in worse Likert ratings. In the baseline condition where image sharpness was not changed, the average Likert rating for all observers is 1.186 out of 4 (1 indicating "no-change"). Likert rating responses significantly depend on the combination of cutoff spatial frequency and post-saccadic duration (p < 0.001). Post-hoc paired t-tests comparing the baseline to the remaining nine conditions show that there are only two conditions where changes in image resolution were perceived similarly to baseline: sf=21.1 cpd at 100 and 200 ms duration (p=0.810 at 100 ms, p=0.393 at 200 ms). These two conditions are also above the acuity thresholds identified in Experiment 1 and therefore expected to be imperceptible (Figure~\ref{fig:exp1_results}). All other conditions are within the detectable regions, and users report noticeable changes relative to baseline at a statistically significant levels (p<0.001).

In addition to analyzing the raw rating responses, we also re-group the ratings into "yes" and "no" responses to derive the probability of detecting a change for each condition (Figure~\ref{fig:exp2}c). Trials in which users responded with "1: no change" are categorized as "no" responses, and all other trials are categorized as "yes" responses. When image sharpness is not manipulated in the baseline condition, users report a perceived change in 17.9\% of trials. At the most aggressive low-pass filtering condition (7.8 cpd with a 500 ms presentation time), users report a visible change in 97.3\% of valid trials. Consistent with the Likert rating results, the probability of reporting a perceived change significantly depends on the cutoff spatial frequency and post-saccadic duration combination (p<0.001). The probability of "yes" responses in the two conditions below the group acuity threshold curve (21.1 cpd at 100 and 200 ms) do not significantly differ from the baseline (p = 0.803 for 21.1 cpd and 100 ms, p = 0.376 for 21.1 cpd and 200 ms; Figure~\ref{fig:exp2}c). For all other conditions, users detect change in low-pass filtered images at a probability significantly higher than that of the baseline (p<0.001). 

Overall the results of Experiment~2 support the generalizability of our low-level psychophysical results from Experiment~1: As shown in Figure~\ref{fig:exp2}b-c, saccade-contingent rendering using parameter points located below our threshold curve were more likely to be detected relative to baseline and also received lower image quality ratings by users, while points above the curve were not significantly different from baseline. In the next experiment, we present a complimentary study and implement saccade-contingent rendering in a headset to evaluate our method in a high-resolution VR headset.

	\section{Experiment 3: In-Headset Validation}
In our last study, we address the tradeoffs made in Experiment 2 and present an ecologically valid, in-headset evaluation of saccade-contingent rendering. We require a headset with low-latency saccade detection, headtracking, and >40 ppd display resolution (Section~\ref{sec:bandwidth}) to serve as a viable testbed for saccade-contingent rendering. In order to meet these requirements, and to completely eliminate the possibility of reaching display resolution-limited image quality, we use a custom-built 90 ppd headset with 30 ms eye movement to photon latency as our test vehicle for this study.

\subsection{Ultra-High Resolution VR Headset}
We utilize a Meta Quest 2 as a pre-existing tracking platform, and retrofit it with PC-driven displays. Two 2880$\times$2880, 3.2" transmissive color LCD displays with an LED backlight were combined with custom-designed viewing optics to achieve 90 ppd over a ~33$^\circ$ field of view with a 1.3 m virtual image distance. The optics are made from high-index glass to achieve diffraction-limited performance and the final design has a focal length of 102 mm. Saccade detection is achieved with an integrated binocular XR eye tracking platform from Tobii (Tobii AB, Sweden). The eye tracker has a potential sampling frequency of 240 Hz and is based on Tobii's latest-generation off-axis (direct to eye) solution for VR and AR optical designs. We measure eye movement to photon latency in our system by monitoring an infrared light with the eye tracking system and render to the display on signal detection. The average measured "eye movement" to photon latency in our headset is 30~ms with maximum values of 50~ms.

\begin{figure}[t]
	\includegraphics[width=.45\textwidth]{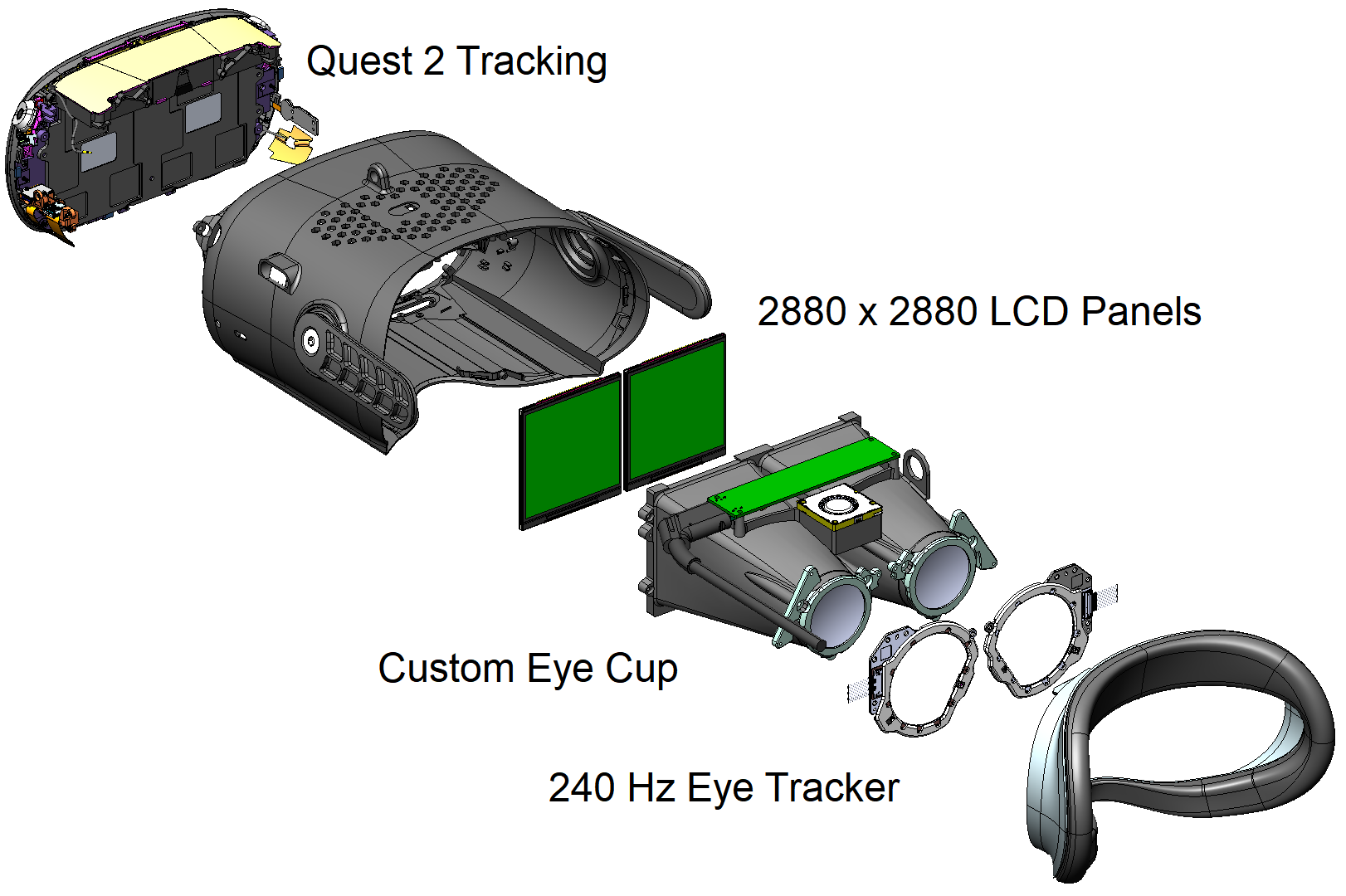}
	\caption{Exploded view of our 90 ppd HMD used to run Experiment 3. This pixel density is achieved by significantly reducing the HMD field of view. This trade-off is made to assess the perceptual consequences of downsampling artifacts from saccade-contingent rendering in a scenario limited by human visual acuity rather than display pixel size. We note that analysis from Section~\ref{sec:bandwidth} identifies potential bitrate savings of 80\% at 90 ppd.}
\end{figure}

\subsection{Rendering Pipeline}
We modified an OpenXR implementation to enable saccade\hyp{}contingent rendering emulation across a wide library of VR content. This allowed us to evaluate the interactions of saccade-contingent rendering within a fully-functioning VR headset software stack. 

\begin{figure*}[t]
	\includegraphics[width=.4\textwidth]{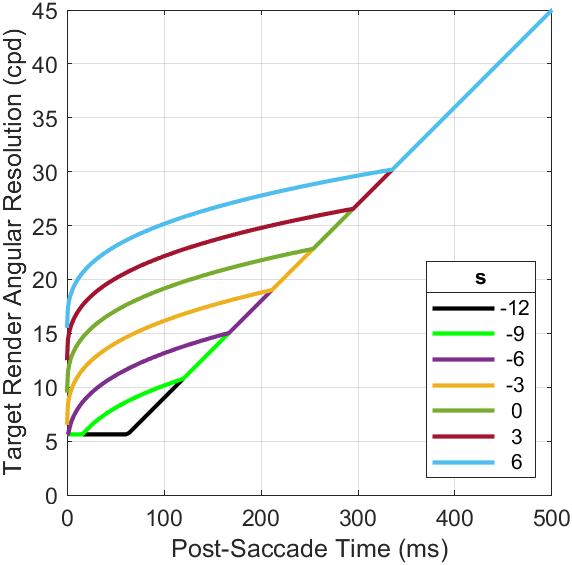}
	\caption{Saccade-contignent rendering downsampling curves evaluated in Experiment~3.}
    \label{fig:exp3curve}
\end{figure*}

Our implementation requests VR applications to render at 1.5x native display resolution, but otherwise does not modify the rendering pipeline until the rendered frame is intercepted at the final stage in the frame compositor. In an additional step we downsample the 1.5x resolution image to the target sub-native resolution using a Gaussian band-pass filter, followed by upsampling using the standard bilinear filtering in the compositor. The downsampling target resolution is guided by Equation~\ref{eqn:sf_curve} where target visual acuity thresholds in cycles per degree are converted to render resolution according to Nyquist theorem (cpd $\times$ 2). For this user study we modify the equation to reach native resolution (45 cpd, 90 ppd) by 500 ms with the introduction of a linear ramp, limiting downsampling to 8$\times$, and introduce an adjustable spatial acuity offset to enable perceptual tuning of the saccade-contingent downsampling curve:
\begin{equation}\label{eqn:exp3}
    sf(t) = \max \left \{ (1.9469t^{0.3475}+9.5062 + s), \frac{t\times45}{500}, \frac{90}{8\times2} \right \}
\end{equation} A family of curves modified by different values of $s$, along with the effect of the linear ramp is shown in Figure~\ref{fig:exp3curve}.



\subsection{User Study}
\begin{figure*}[t]
	\includegraphics[width=.8\textwidth]{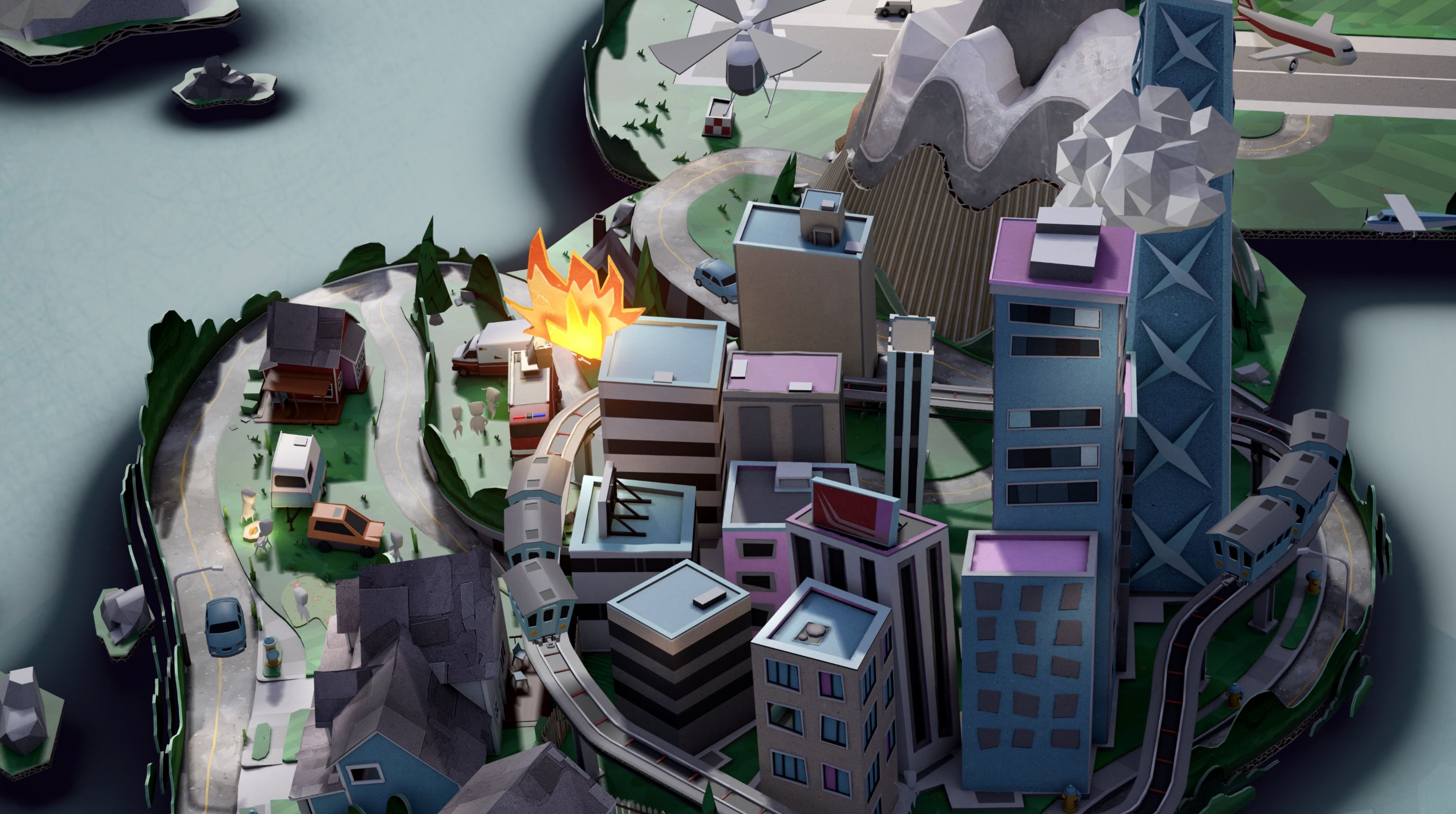}
	\caption{Paper Town scene used to evaluate saccade-contingent rendering in Experiment 3.}
    \label{fig:papertown}
\end{figure*}

In our final user study 16 observers (14 male, 2 female) evaluate image quality using a four point Likert scale (1 undetectable, 2 not annoying, 3 annoying, 4 unacceptable) while saccade-contingent rendering is applied with different values of $s$. Users are instructed to visually explore the Paper Town scene (Figure~\ref{fig:papertown}) from Oculus Dreamdeck (Meta, Menlo Park, CA, USA). Users can make head movements, but are asked to avoid fast head turns to avoid potential false saccade detection from high-velocity vestibulo-ocular reflex eye movements. Users evaluate five values of $s$ ranging from zero (default curve from Experiment~1) to -12. All but three users rated artifacts from saccade-contingent rendering while viewing the default curve as "undetectable." The three users that rated the default downsampling curve as "detectable, but not annoying" are shown positive $s$ values to reduce the amount of downsampling until they report an "undetectable" rating ($s$ values >0 are rated "undetectable" for the other observers). The average reported Likert score across each value of $s$ are shown in Table~\ref{table:exp3}.

\begin{table}[H]
\label{tabel:exp3}
\begin{center}
\begin{tabular}{ |c|c|c|c|c|c|c|c| }
    \hline
    $s$ & -12 & -9 & -6 & -3 & 0 & 3 & 6\\
    \hline
     Mean Rating  & 3.9  & 3.3& 2.2  & 1.4  & 1.2 & 1.1 & 1 \\
    $\sigma$ & $\pm 0.3$  & $\pm 0.7$ & $\pm 0.7$ & $\pm 0.6$ & $\pm 0.4$ & $\pm 0.3$ & $\pm 0$\\
    \hline
\end{tabular}
\caption{Mean Likert ratings for different values of $s$ in Experiment~3. Ratings from one to four correspond, respectively, to undetectable; detectable, but not annoying; detectable and annoying; and unacceptable.}
\label{table:exp3}
\end{center}
\vspace{-6mm}
\end{table}

Our final user study supports previous results from Experiments~1 and 2; all observers rate the subjective, in-headset experience as "detectable, but not annoying" at worst, with 13/16 reporting a rating of "undetectable." This final study provides additional, complementary validation of saccade-contingent rendering in a setting that is as close as possible to "real-world" conditions with users experiencing images affected by acutal HMD head tracking, eye tracking, and rendering pipelines. The consistent findings across our studies provides strong validation for the feasibility of saccade-contingent rendering. Additionally, our saccade-contingent rendering pipeline simply emulates rendering at sub-native display resolution, which means our method can be easily implemented without major architectural revisions in existing HMD rendering pipelines.	
	\section{Discussion}
We present three user studies that comprehensively characterize reductions in human spatial acuity after saccades. We also demonstrate potential compute and data savings by leveraging this phenomenon to render at lower resolution for several hundred milliseconds following every saccade. Under natural viewing conditions this algorithm can yield bandwidth savings in excess of 18\% for 45 ppd displays and 53\% for 60 ppd displays. We conclude with potential limitations of our approach and other considerations.

\paragraph{Interactions with Image Content}
The spectrum power of natural images typically exhibit a $1/f^2$ frequency falloff, meaning that energy of high spatial frequency information is exponentially less than that of low spatial frequency information~\cite{tolhurst1992amplitude,ruderman1994statistics}. Rendering images at lower resolution effectively cuts off the amount of high spatial frequency energy that is presented to the user. However, some types of image content diverge significantly from $1/f^2$ frequency characteristics. Text, for example, exhibits a power spectrum that is much flatter in the frequency domain and therefore rendering at reduced resolution disproportionately affects its frequency spectrum compared to a natural image. Conversely, low-contrast images with less high spatial frequency content can be more aggressively downsampled. More specific parameter tuning, or even a dynamic content-aware implementation, can be studied to determine the best balance between bitrate savings and use-case-driven tolerance for visible downsampling artifacts. 

\paragraph{Integration with Existing Methods}
Post-saccadic changes in contrast sensitivity are not currently accounted for in spatiotemporal models used for perceptually\hyp{}optimized rendering. Thus, saccade-contingent rendering could be directly implemented alongside other human contrast sensitivity-optimized rendering techniques without additional refinement. Savings would be reduced if other methods drop the target render resolution below thresholds from Equation~\ref{eqn:sf_curve}, like in foveated rendering where spatial resolution is lowered in the periphery. However, additional joint optimizations are possible. For example, our study in Experiment~1 does not examine eccentricity and similar ratios of post-saccadic acuity reduction have been identified at four and eight degrees of visual eccentricity \cite{li2021post}.

\paragraph{Event-Sensor Based Eye Tracking}
Our proposed saccade\hyp{}contingent rendering takes advantage of the fact that visual acuity over the \textit{entire} visual field is reduced after a saccade. Thus, unlike foveated rendering techniques, the spatial accuracy of the eye tracking system is irrelevant, and only low-latency, accurate saccade detection is necessary to implement our technique effectively. Such requirements could be achieved using non-traditional eye tracking solutions, such as event sensors, to achieve lower power consumption and more compact form factor~\cite{Angelopoulos2021}.

\paragraph{Varifocal Saccade-Contingent Rendering}
In our psychophysical study, saccades were made to the same distance in depth. Saccades made in natural viewing frequently coincide with gaze changes to different egocentric distances \cite{gibaldi2019binocular}. Such eye movements require the eye to accommodate, a process which takes several hundred milliseconds \cite{lockhart2010effects, bharadwaj2005acceleration}. Defocus blur will affect the retinal images until the eye is properly focused, and during this time it is unlikely that microsaccades and ocular drifts will improve visual acuity. It is plausible that saccade-contingent rendering could be extended for an even longer duration to account for this in a varifocal VR HMD \cite{padmanaban2017optimizing}. High-resolution displays are most beneficial when paired with varifocal capabilities \cite{zhao2023retinal}, and the application of saccade-contingent rendering could offer even more significant savings when users make combined saccade and accommodative eye movements in varifocal HMDs. To identify accurate perceptual thresholds in this scenario, the protocol in Experiment~1 should be repeated with a saccades to different viewing distances.

\section{Acknowledgements}
We thank Joel Hegland and Dave Lindberg for mechnical engineering support, Julia Majors and Douglas Lanman for copy editing, and Shih-Kuang Chu for software assistance.
	\bibliographystyle{ACM-Reference-Format}
	\bibliography{bib}

\end{document}


\title{Saccade-Contingent Rendering}
\subtitle{\Large{Supplementary Material}}
\maketitle
\vspace{1em}

\section{Experiment 1}

\subsection{Additional User Study Details}\label{sec:exp1_onlineanalysis}
The initial fixation circle used in our stimulus is at the center of the display and had a radius of $0.4^\circ$ and was presented on a gray background (7.23 cd/m$^2$). The saccade targets appeared 10.3$^\circ$ to 15.5$^\circ$ away in steps of 1.3$^\circ$. The radius of the target and the distractor Gabor stimuli were $0.7^\circ$, and the distractors were $0.7^\circ$ away from target center. Users were instructed to move their eyes to the center of the target as precisely as possible. The total presentation duration of the stimuli and noise masks were fixed to a combined duration of 550 ms. Gaze position was monitored online to ensure correct fixation and precise eye movements. The start and end of saccades were marked when the gaze first deviated more than 1.9$^\circ$ away from fixation after stimulus onset and when the gaze first landed within 2.9$^\circ$ of the center of the target stimulus location. Trials in which gaze deviated from 1.9$^\circ$ radius of fixation \textit{before} stimulus onset were aborted and repeated at the end of each block. Additionally, we repeated trials when saccades did not land within 2.9$^\circ$ from the center of the target stimulus location. For data analysis a more accurate, velocity-based saccade detection algorithm was used to identify saccade onset and landing (supplementary materials Section~\ref{sec:exp1_eyeana}).

\subsection{Modeling and Threshold Extraction} \label{sec:exp1_aepsych}
We used adaptive sampling in AEPsych ~\cite{owen2021,letham2022} to efficiently select points to test in our two-parameter space - spatial frequency and post-saccadic duration - for data collection and model fitting. In the AEPsych model, the upper and lower bounds for spatial frequency were set to 7.8 and 21.1 cycles per degree (cpd) respectively. The lower and upper bounds for post-saccadic duration were 0 and 500 ms respectively.

When the stimulus's presentation duration was long enough or its spatial frequency low enough, participants could more easily identify its orientation. Vice versa, when the duration was too short, or the spatial frequency too high (or both) participants could not reliably determine the orientation of the target, reporting its orientation at or near chance (50\%). For each presentation duration we identified peak spatial acuity as the highest spatial frequency Gabor stimulus where its orientation was correctly identified 75\% of the time. We extracted the 75\% correct acuity thresholds by re-fitting non-parametric AEPsych models to user responses (Figure~\ref{suppfig:userdataexp1}). The post-saccadic duration values for re-fitting the model were adjusted based on the more precise saccade offset time points obtained through running the offline saccade detection algorithm (Section~\ref{sec:exp1_eyeana}). For modeling and threshold extraction, we combined responses for the 20 different stimulus locations (four corners and five eccentricities). 

\subsection{Offline Eye Movement Analysis} \label{sec:exp1_eyeana}
In addition to the online saccade detection described in Section~\ref{sec:exp1_onlineanalysis}, we also performed offline analysis using an established algorithm for saccade detection \cite{EngbertEtAl2006}. Saccades were detected based on their velocity distribution using a moving average over 20 eye position samples. Saccade onsets and offsets were detected based on when the velocity exceeded or fell below the median of the moving average by three standard deviations for at least 20 ms. This offline analysis takes the whole distribution of eye position samples and velocity into consideration, and allows for a more detailed evaluation of both fixational and saccadic eye movements. Post-saccadic stimulus presentation durations were determined based on the output of the offline saccade detection algorithm and used in place of the stimulus presentation durations generated online for AEPsych model fitting described in Section~\ref{sec:exp1_aepsych}. 

\begin{figure}[h]
	\includegraphics[width=.975\textwidth]{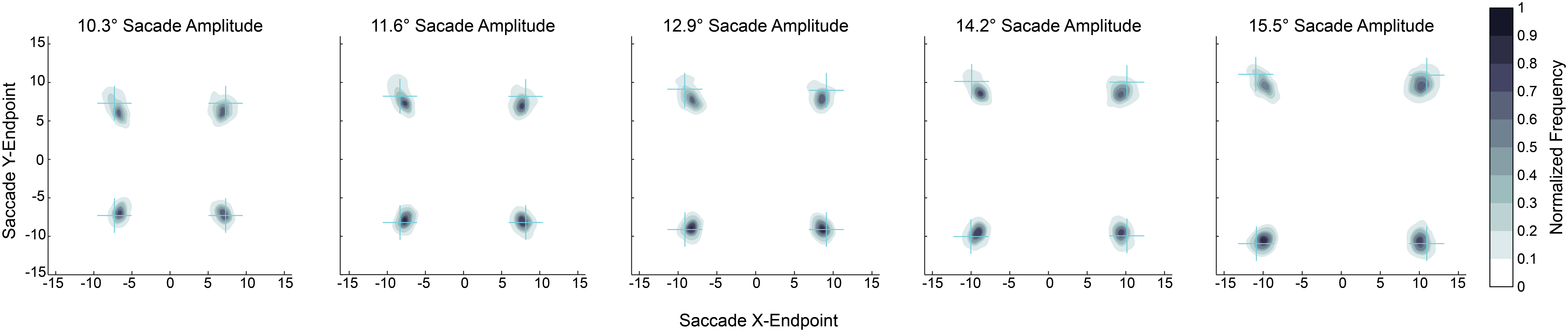}
	\caption{Normalized saccade endpoint frequency maps averaged across participants as a function of saccade direction for five different saccade amplitude trials. Crosses indicate the 2.9$^\circ$ saccade target region.} 
	\label{fig:suppl_exp1_sacLand}
\end{figure}

\subsection{Experiment 1 Psychophysical Data} \label{suppfig:userdataexp1}
\begin{figure}[h]
	\includegraphics[width=\textwidth]{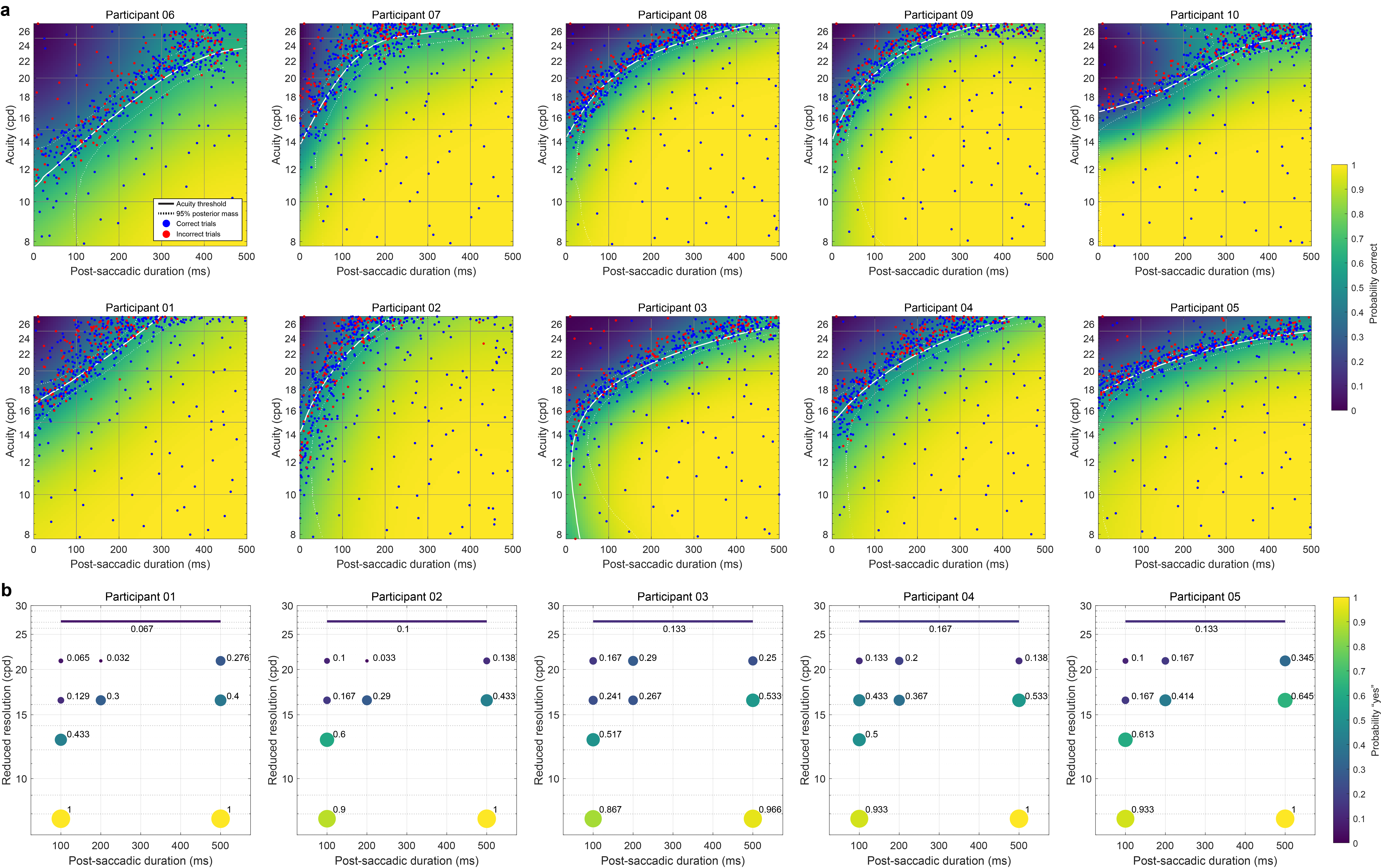}
	\caption{Data of individual observers. (a)  Visual acuity thresholds (Exp1) of 5 observers who did not participate in Experiment~2. (b) Visual acuity thresholds of users (Experiment~1) and probability of change detection (Experiment~2) of five individual observers who participated in both of the experiments. The individual in the far right figure is the example observer shown in Fig~1b.}
	\label{fig:suppl_aepsych}
\end{figure}

\newpage
\section{Experiment 2}
\subsection{Image Processing}
Each of the 30 images used in this study was cropped to 4000 $\times $ 4000 pixels, down-sampled to 2000 $\times $ 2000, and low-pass filtered to 7.8, 12.8, 16.4, 21.1, or 27.1 cpd, depending on the trial condition (Section~4.1.
A Butterworth filter was used for lowpass filtering to avoid ringing artifacts, and the filter is applied in the Fourier domain separately for each color channel at the specified spatial frequency cutoff values.

\subsection{Additional Data}
\begin{figure}[h]
	\includegraphics[width=.975\textwidth]{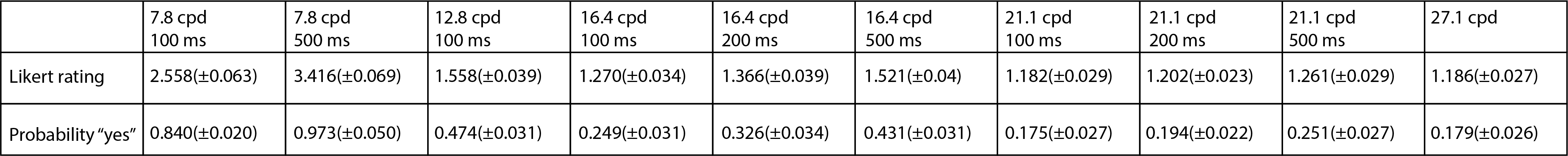}
	\caption{Exp2: Mean responses (Likert ratings and probability of change detection) and $\pm1$ standard error of the mean as a function of reduced resolution and post-saccadic duration}
	\label{fig:suppl_exp2_table}
\end{figure}

\begin{figure}[h]
	\includegraphics[width=.975\textwidth]{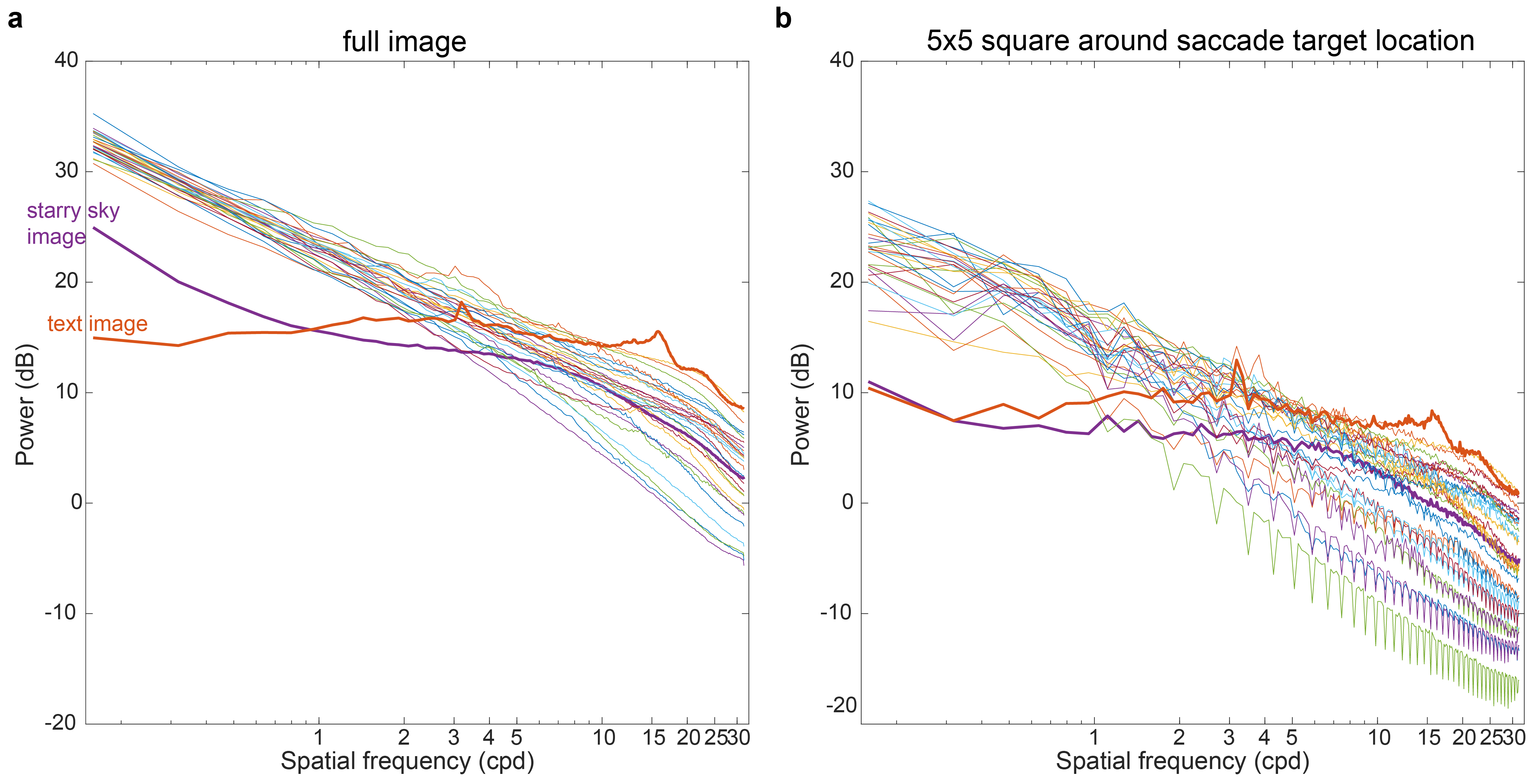}
	\caption{Power spectrum of images used in Experiment 2, of (a) the full image or (b) the 5 $\times $ 5 deg$^2$ area around the saccade target location of each image. Each line indicates one image. The text image has a flatter power spectrum across spatial frequencies than other natural images.} 
	\label{fig:suppl_fft}
\end{figure}

\begin{figure}[h]
	\includegraphics[width=.75\textwidth]{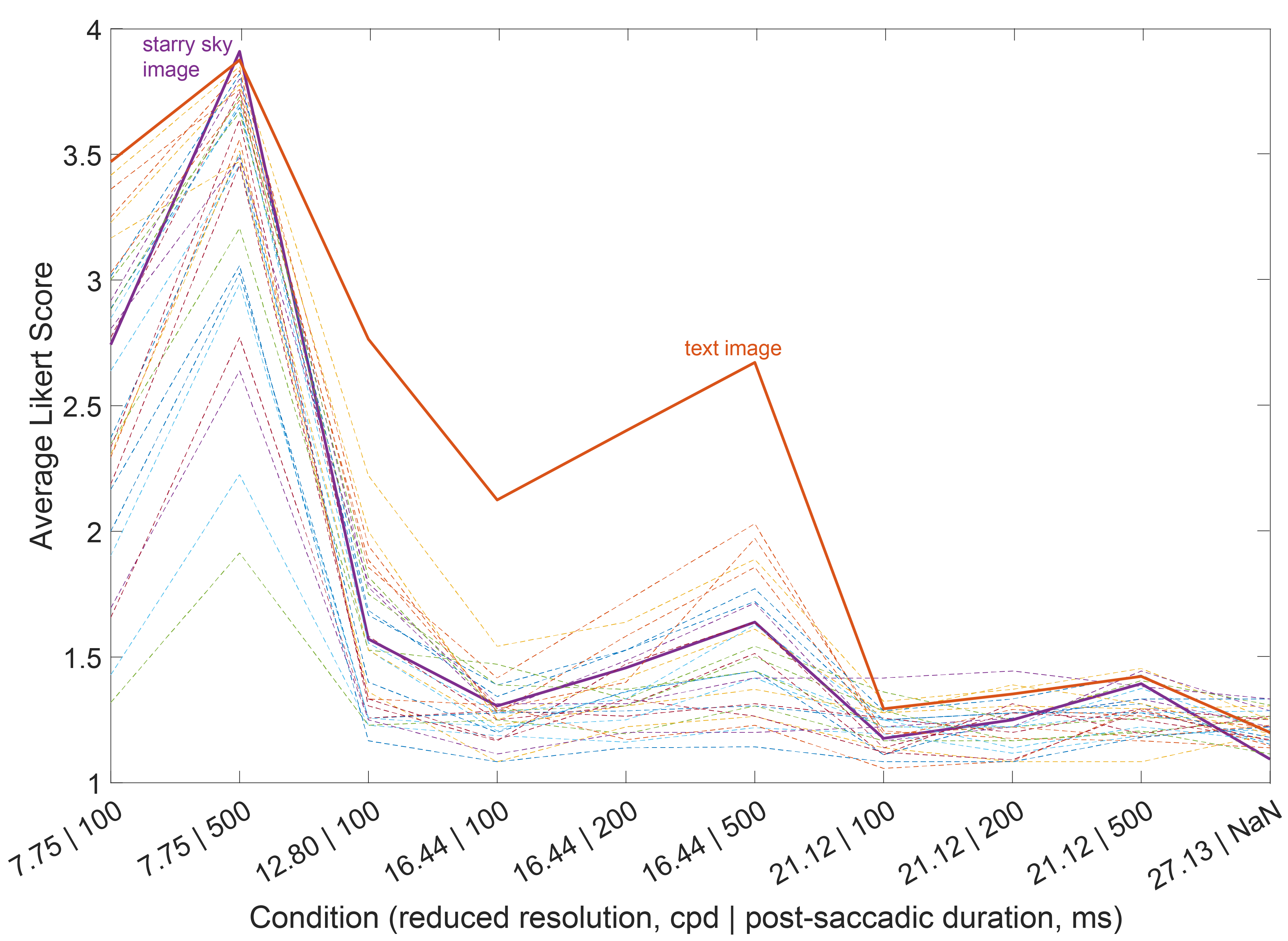}
	\caption{Average Likert scores per image under each condition. Each line indicates the average Likert scores of one image. }
	\label{fig:suppl_probPerImg}
\end{figure}

\newpage

\subsection{Image sources} \label{sec:suppl_img}
\begin{enumerate}
\item L1440646, Richard Lazazara, \url{https://www.flickr.com/photos/shankargallery/52575102416}

\item tube lavande, Rafael Wagner, \url{https://www.flickr.com/photos/125877475@N06/53099508605}

\item untitled, Uwe Brodrecht, \url{https://www.flickr.com/photos/uwebrodrecht/53102181732/}

\item Vickers Vixcount 837 XT575, Michael Gaylard,  \url{https://www.flickr.com/photos/mgaylard/53102648793}

\item untitled, La Grande Oregon Economic Development,

\url{https://www.flickr.com/photos/lagrandeeconomicdevelopment/53103112928/}

\item untitled, La Grande Oregon Economic Development,

\url{https://www.flickr.com/photos/lagrandeeconomicdevelopment/53103246498/}

\item untitled, Uwe Brodrecht, \url{https://www.flickr.com/photos/uwebrodrecht/53103282793/}

\item untitled, Alexandr Istomin, \url{https://unsplash.com/photos/YTALkNd4rO0}

\item untitled, Rick Barrett, \url{https://unsplash.com/photos/\_7w0FPz5mxQ}

\item untitled, Andrea Davis,  \url{https://unsplash.com/photos/SoRlz-tnWUM}

\item untitled, Anita Austvika, \url{https://unsplash.com/photos/onhSSKszNCg}
 
 \item untitled, Boglárka Salamon,  \url{https://unsplash.com/photos/dYfPZk5o6LI}

\item untitled, Clement Proust,  \url{https://unsplash.com/photos/TaLgSVSOTQE}

\item untitled, D5 Render, \url{https://unsplash.com/photos/O3LR3ZI5Ams}

\item untitled, Daniel J. Schwarz, \url{https://unsplash.com/photos/oARPb2dEOtQ}

\item untitled, Danist Soh, \url{https://unsplash.com/photos/vS1QLaUKRHg}

\item untitled, Gemali Martinez, \url{https://unsplash.com/photos/-hUq3ah7szI}

\item untitled, Jasper Garratt, \url{https://unsplash.com/photos/cjHRI1OwKbk}

\item untitled, Jiran Family, \url{https://unsplash.com/photos/SpLbbq5yAR4}

\item untitled, Justin Wolff, \url{https://unsplash.com/photos/vx4Nuvb40jc}

\item untitled, Loewe Technology, \url{https://unsplash.com/photos/04fQqBaJwcw}

\item untitled, Loewe Technology, \url{https://unsplash.com/photos/VtwpO_4Txm8}

\item untitled, Logan Weaver, \url{https://unsplash.com/photos/qz6k5OWoUL0}

\item untitled, Rhamely, \url{https://unsplash.com/photos/Y1ep8QjOmBQ}
  
\item untitled, Ryan Vargas, \url{https://unsplash.com/photos/3kaxEQI0ruc}

\item untitled, Timothy Chambers, \url{https://unsplash.com/photos/8H7L5VdE81c}
  
\item untitled, Timo Volz, \url{https://unsplash.com/photos/qkA80ZzX2CM}
  
\item untitled, Tobias Reich, \url{https://unsplash.com/photos/XOCm_iufXEk}

\item untitled, Yuriy Vertikov, \url{https://unsplash.com/photos/xRnK7DtTI4I}

\end{enumerate}



\bibliographystyle{ACM-Reference-Format}
\bibliography{bib}